\documentclass[11pt,twoside]{article}
\usepackage{asp2010}
\usepackage{color}
\resetcounters

\begin{document}

\title{Calibrating the role of TP-AGB stars in the cosmic matter cycle}
\author{Paola~Marigo
\affil{Dipartimento di Fisica e Astronomia, Universit\`a di Padova, Italy}
}
\begin{abstract}
In the last ten years three main facts about the thermally pulsing asymptotic giant branch 
(TP-AGB) have become evident: 1) the modelling of the  TP-AGB phase is critical 
for the derivation of basic galaxy properties (e.g.\ mass and age) up to high redshift,
with consequent cosmological implications;
2) current TP-AGB calibrations based on Magellanic Cloud (MC) clusters come out not to
work properly for other external galaxies, yielding a likely TP-AGB overestimation; 
3) the significance of the TP-AGB contribution in galaxies, hence their derived properties,  
are strongly debated, with conflicting claims in favour of either a {\em heavy} or a 
{\em light} TP-AGB.
The only way out of this condition of persisting uncertainty is to perform 
a reliable calibration of the TP-AGB phase as a function of the star's initial mass (hence age)
over a wide range of metallicity, from very low to super-solar values.
In this context, I will  review recent advancements and ongoing efforts  
towards a physically-sound TP-AGB calibration that,  moving beyond the 
classical use of the MC clusters, combines increasingly refined TP-AGB 
stellar models  with exceptionally high-quality data 
for resolved TP-AGB stars in nearby galaxies.
Preliminary results indicate that
a sort of ``{\em TP-AGB island}'' emerges in the age-metallicity plane, 
where the contribution of these stars is especially developed, 
embracing preferentially solar- and MC-like 
metallicities, and intermediate ages ($\sim$ few Gyr).
\end{abstract}
\section{Broad context: 
the TP-AGB issue in galaxy models}
\label{sect_intro}
It has been known for long time that,
owing to their high intrinsic brightness, TP-AGB stars contribute
significantly to the total luminosity of single-burst stellar
populations, reaching a maximum of about 40\% at ages from 1 to
3 Gyr \citep{Frogel_etal90}, and accounting for most of
the infrared-bright objects in resolved galaxies, as clearly
demonstrated in the Magellanic Clouds 
\citep[MC; e.g.,][]{Bolatto_etal07, Blum_etal06, NikolaevWeinberg_00}.
\begin{figure*}
\begin{center}
\begin{minipage}{0.39\hsize}
\resizebox{1.0\hsize}{!}{\includegraphics{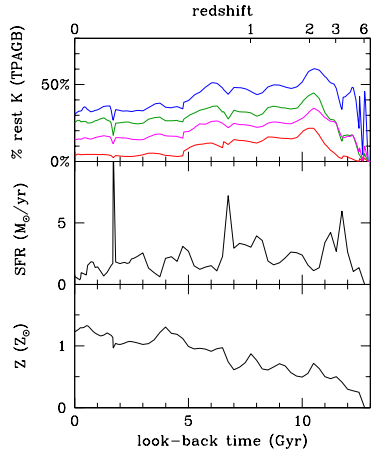}}
\end{minipage}
\hfill
 \begin{minipage}{0.60\hsize}
\caption{Predicted contribution by the TP-AGB to
the (rest-frame) K-band flux, the cosmic star formation rate (SFR), and the 
metal enrichment ($Z$) as a function of the look-back time, derived from SPS models 
applied to the Millennium Cosmological Simulation \citep{Springel_etal05}.
The different curves correspond to different
prescriptions of the TP-AGB phase, mainly related to various
efficiencies of mass loss and third dredge-up.
It also shows
how the TP-AGB contribution is boosted after each major episode of star formation
(i.e. in the post-starburst phase). Courtesy of S. Charlot and G. Bruzual.}
\label{fig_uncertk}
\end{minipage}
\end{center}
\end{figure*}

However, the high influence of TP-AGB stars in stellar population
synthesis (SPS) models of galaxies was  recognized 
only after \citet[][herefater M05]{Maraston_05} pointed out that they can dramatically
alter the mass-to-light ratio for 0.5 -- 2 Gyr old stellar
populations, hence affecting the determination of stellar masses and ages
of high-redshift galaxies by factors of 2 or more. 

While the importance of the TP-AGB phase is now universally acknowledged, since 
M05 a number of {\em conflicting results} have been produced
about the {\em overall impact of TP-AGB stars}.
In this context, many recent papers have focused on comparing the
performances of two popular SPS models that differ radically in the technique adopted to
include the TP-AGB phase.  \citet[][hereafter BC03]{BruzualCharlot_03} 
use sets of stellar evolutionary tracks, while in the M05 models the TP-AGB contribution is
described by the integrated nuclear fuel (emitted light), bypassing the details of stellar
evolution.
The appropriateness of M05 (favouring  a heavy TP-AGB) and BC03 
(characterized by a light TP-AGB) models is a matter of lively debate, yielding
discordant claims.
For instance, while M05 appears to overpredict the near-infrared part of the 
spectral energy distributions of post-starburst galaxies at moderate-to-high redshifts 
\citep{Kriek_etal10, Zibetti_etal13} and BC03 models are better performing, 
the observed HCN spectral features in the nuclear regions of AGNs are 
in closer agreement with M05 models \citep[see also Riffel's contribution, 
this conference]{Riffel_etal07}.

\begin{figure*}
\begin{center}
\begin{minipage}{0.49\hsize}
\resizebox{0.9\hsize}{!}{\includegraphics{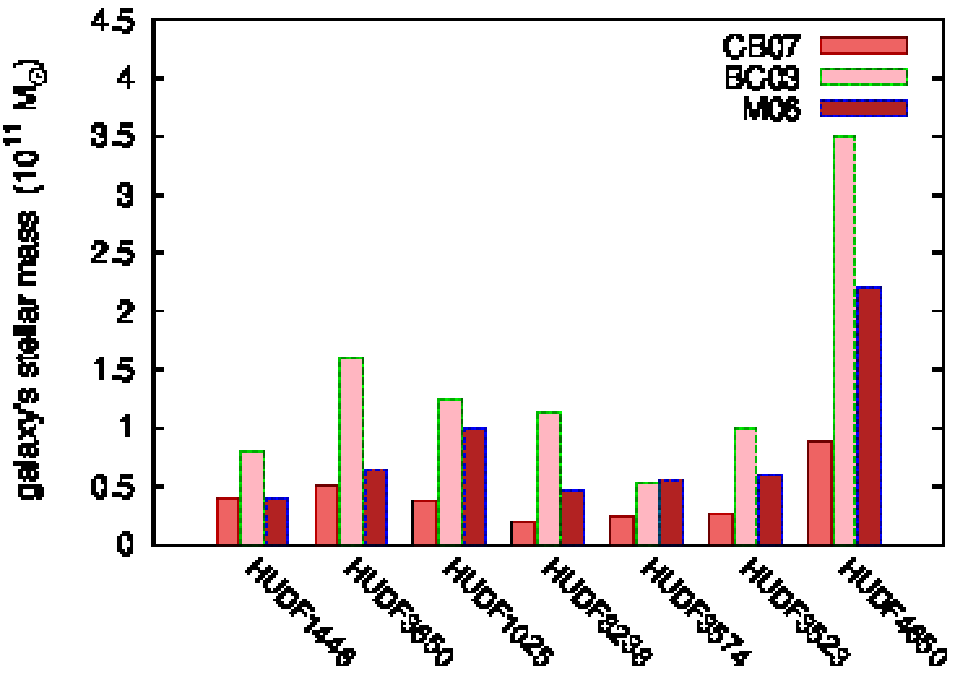}}
\end{minipage}
\hfill
\begin{minipage}{0.49\hsize}
\resizebox{0.9\hsize}{!}{\includegraphics{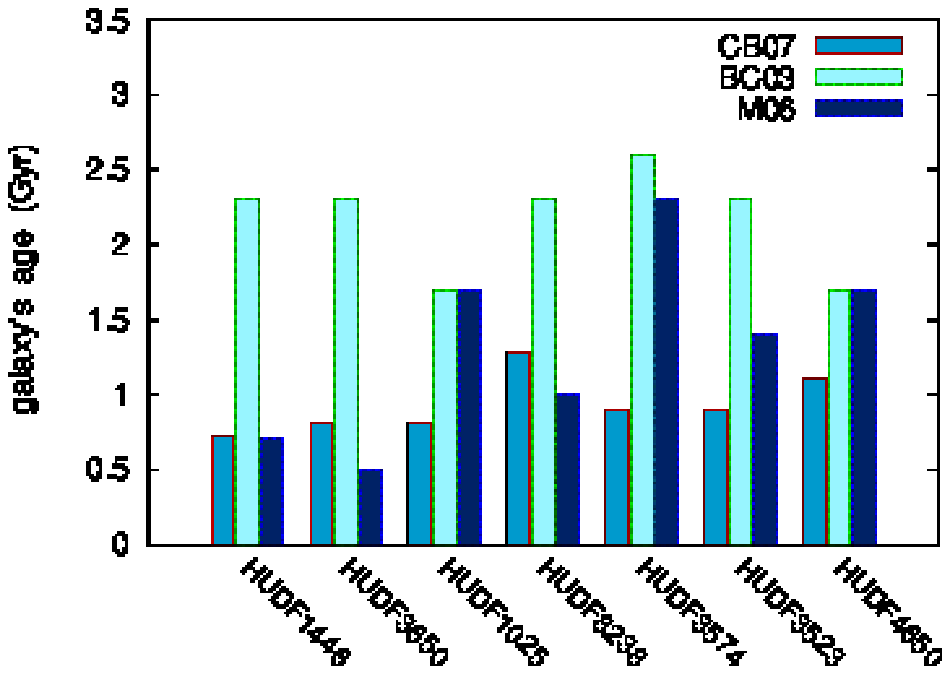}}
\end{minipage}
\caption{An example of the current uncertainties 
in the mean ages and stellar masses of galaxies (up to a factor of $\sim$ 2-3),
due to the propagation of the uncertainties in the underlying TP-AGB models. 
The bars show the estimates for a sample of seven intermediate-redshift galaxies in the 
Hubble Ultra-Deep Field (HUDF) derived with three different SPS models that mostly 
differ in the treatment of the TP-AGB:  \citet[][BC03]{BruzualCharlot_03}, 
\citet[][M06]{Maraston_etal06}, \citet[][CB07]{Bruzual_07} 
based on the TP-AGB tracks of \citet{MarigoGirardi_07}.}
\label{fig_uncert_massage}
\end{center}
\end{figure*}

What strikingly emerges is that
the uncertainties intrinsic to population synthesis models   -- and especially those related 
to the TP-AGB contribution -- often dominate over the observational errors when drawing
galaxy properties across cosmic times, impinging dramatically on 
the derivation of the stellar mass-to-light-ratios, 
masses and ages from the integrated light of galaxies 
\citep[][see also Figs.\,\ref{fig_uncertk}-\ref{fig_uncert_massage}]
{Conroy_13, Taylor_etal11, Zibetti_etal09, Eminian_etal08,  Bruzual_07, Maraston_etal06}.

The inescapable bottom line is that a decisive step forward in the
calibration of the TP-AGB phase is urgently needed!
As we will show, this challenging goal
may be only achieved with TP-AGB stellar evolution models, able to predict individual 
physical properties to be compared with observations of resolved stars. 
The potential contribution of other approaches, e.g.\ those based on the integrated 
TP-AGB fuel, is much weaker as they cannot, by construction, 
be tested against the wealth of information about resolved 
TP-AGB stars available nowadays (Sect.~\ref{sect_obs}).
\section{Magellanic Cloud clusters as TP-AGB calibrators}
\subsection{The classical approaches and insidious problems}
Historically,  the calibration of the TP-AGB as a function of the age is based 
on the globular star clusters  in the Magellanic Clouds (MCs), relying on the star counts, 
integrated fluxes and spectral classification.
The pioneer work in this field was carried out by \citet{Frogel_etal90}, 
who derived the fractional luminosity contributed by TP-AGB stars as well 
as the luminosity functions of M and C stars  as a function of SWB type, which is a proxy 
for the age \citep{Searle_etal80}.
Since that work, following studies employed the MC cluster data of AGB stars 
for calibration purposes, either using the measured integrated luminosities or broad-band 
visual and near-IR colors to estimate the nuclear fuel burnt during the TP-AGB phase
\citep{Maraston_05, Noel_etal13}, or dealing with the direct star counts to derive 
the TP-AGB lifetimes with the aid of stellar tracks and isochrones 
\citep{CharlotBruzual_91, GirardiMarigo_07}.

Despite these calibration efforts based on different techniques, when present-day  
TP-AGB models are applied to other external galaxies  
they overestimate, to various extents,
the TP-AGB contributions in integrated spectra of galaxies or star counts. 
For instance, M05 models show an excess of IR flux that is not 
observed in post-starburst galaxies \citep{Kriek_etal10, Zibetti_etal13}. 
On the other hand, \citet{MarigoGirardi_07} 
tracks predict, on average, $40\%$ more TP-AGB stars than counted in
a sample of nearby galaxies observed with ANGST, 
which translates in a factor of $\sim 2$ in the integrated near-IR flux \citep{Melbourne_etal12}.

The question that arises is, therefore:
Why TP-AGB models calibrated on MC clusters are not equally adequate  
for other galaxies, even with metallicities comparable to the MCs? 

\begin{figure}
\begin{center}
\resizebox{0.44\hsize}{!}{\includegraphics{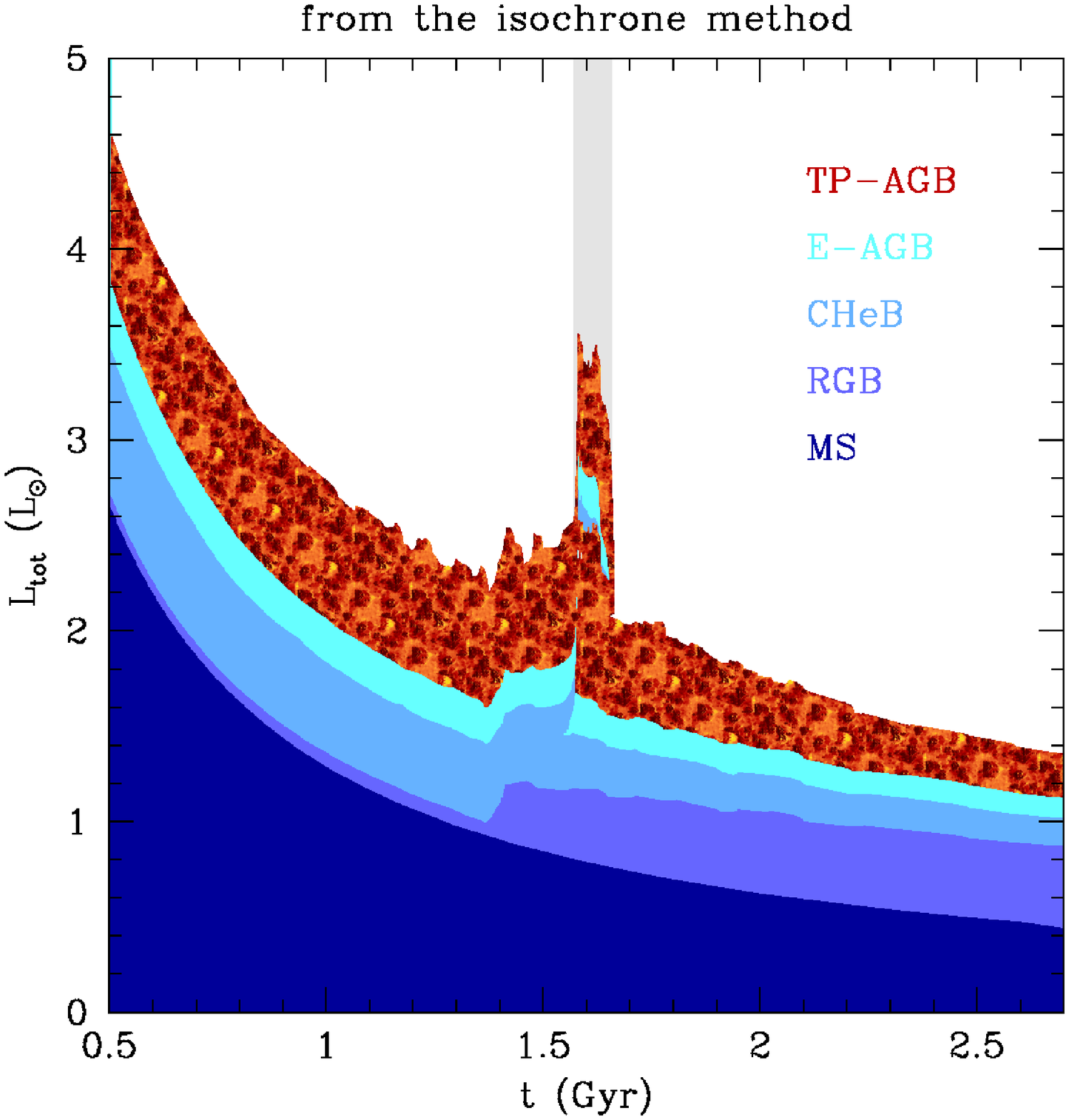}}
\caption{Evolution of the integrated bolometric 
luminosity of a simple stellar population as derived from detailed stellar 
tracks and isochrones,  
as a function of age \citep{Bressan_etal12, Marigo_etal13, Girardi_etal13}.
From bottom to top the sequence of different evolutionary phases is 
shown: main sequence (MS), 
red giant branch (RGB), core helium-burning (CHeB), and Early-AGB (E-AGB). 
The brightest region, filled with wavy pattern, refers to the TP-AGB contribution.  
The AGB-boosting period at ages  $\sim$1.6 Gyr is marked by the shaded 
vertical bar. }
\label{fig_boost}
\end{center}
\end{figure}

Recently, \citet{Girardi_etal13} have pointed out a specific aspect,
related to the physics of stellar interiors, that is likely the main cause of 
this conundrum. As soon as stellar
populations attains the ages at which red giant branch stars first arise, 
an abrupt increase in the lifetime
of the core He-burning phase causes a transitory boost in the production 
rate of the later evolutionary phases,
including the TP-AGB. For a time interval of about 0.1 Gyr, triple TP-AGB branches 
grow at somewhat different initial
masses, making their frequency and contribution 
to the integrated luminosity of the stellar population to raise 
by a factor of $\simeq 2$ (see Fig.~\ref{fig_boost}). 
The boost takes place for turn-off masses of $\simeq 1.75 M_{\odot}$, 
just in vicinity of the predicted peak in the TP-AGB lifetimes for MC metallicities, 
and for ages of $\simeq 1.6$ Gyr (see Fig.~\ref{fig_tauall}). 
Coincidently, this relatively narrow age
interval contains the few very massive MC clusters 
where most of the TP-AGB stars, used to constrain stellar evolution and SPS models, 
are found.
As a consequence, the expected boosting of TP-AGB stars in intermediate-age MC clusters 
may account for the excess of TP-AGB in current models.
 
Two main implications can be drawn:  i) all classical estimates on 
the relative role of TP-AGB stars to the integrated light of intermediate-age 
stellar populations are likely biased 
towards too high values \citep[including][]{Maraston_05, Marigo_etal08}, and ii) 
TP-AGB star populations in intermediate-age MC clusters await to be carefully 
revised, promisingly with the aid of sets of stellar evolutionary 
tracks and isochrones calculated with the level of detail necessary 
to reveal the TP-AGB boosting.

\subsection{Enhancing the classical TP-AGB calibration based on MC clusters}
Recently, the classical use of MC clusters, commonly based on photometry and star 
counts,  has been expanded to include other key properties  
of TP-AGB stars, namely: pulsation and nucleosynthesis. 
We have now information about the pulsation periods and pulsational masses 
of TP-AGB stars and their location on different period-luminosity sequences
(that are likely related to different pulsation modes), as well as on surface chemical abundances, 
in particular the C/O ratio, the $^{12}$C/$^{13}$C isotopic ratio, 
and other elements affected by the TP-AGB nucleosynthesis (e.g., F, Li)
\citep{Kamath_etal12,  Kamath_etal10, Lederer_etal09, 
Lebzelter_etal14, Lebzelter_etal08, LebzelterWood_11,  LebzelterWood_07, Maceroni_etal02}.
 
The challenge that AGB models have to face is to reproduce various
observables at the same time, i.e the giant branch temperatures,
the oxygen to carbon transition luminosity (when C/O overcomes unity due
to the third dredge-up), the AGB-tip luminosity, the period-luminosity relations,  
and the observed elemental abundances.

Despite severe difficulties still unsolved,  these studies have 
clearly shown that, with the aid of targeted observations coupled to 
detailed TP-AGB nucleosynthesis calculations and a parametrized description 
of the convective boundaries, 
it is possible to derive important constraints on crucial but poorly
known quantities, such as the extent of the envelope overshoot, 
the depth of the partially-mixed zone, the intershell composition,  
the minimum core mass and the efficiency of the third dredge-up.

These all-round approaches are promising and deserve 
to be developed further, hopefully also in combination with population synthesis studies.
\section{The need to move beyond the MC clusters: 
wide age-metallicity sampling and characterization}
\label{sect_obs}
\begin{figure}
\begin{center}
\resizebox{0.52\hsize}{!}{\includegraphics{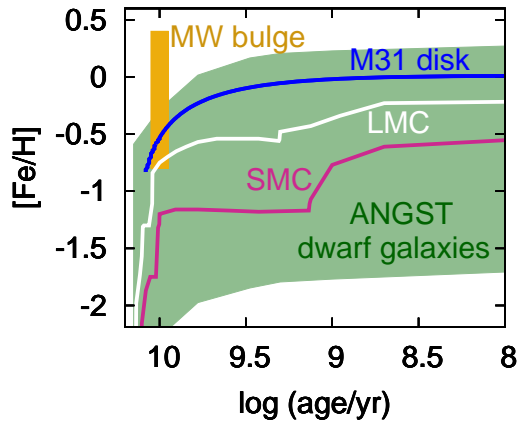}}
\caption{The relevant area in the age--metallicity diagram 
that needs to be covered for a reliable 
TP-AGB calibration. Notice that the inclusion of M31 (PHAT data), 
 of ANGST dwarf galaxies, and of the Galactic bulge (OGLE, WISE) 
largely increases the sampling of this plane, 
compared to previous calibrations \citep{Maraston_05, Marigo_etal08} 
which were based only on the Magellanic Clouds.}
\label{fig_amr}
\end{center}
\end{figure}

Though the MC clusters represent a key benchmark to test and calibrate
the TP-AGB models, at the same time they are affected by two severe 
limitations, namely: the low-number statistics of TP-AGB stars with associated
large Poisson fluctuations, and the  narrow sampling of the
age-metallicity plane (see Fig.~\ref{fig_amr}). Moreover, most star
clusters have uncertain ages, and previous analyses do not include the
fact that many of these clusters have multiple age sub-populations and they
cannot be assumed to be single stellar populations 
\citep[][and references therein]{Goudfrooij_etal14}.

With the new generation of large imaging and spectroscopic surveys of
nearby galaxies, the quantity and quality of the data potentially useful for
calibrating TP-AGB models are dramatically increasing.
This data set will make it possible (i) to improve substantially the exploitation of the MC 
data thanks to the detailed and spatially-resolved star formation histories (SFH) 
for fields and clusters, and (ii) to
extend largely the metallicity range of the calibration, as shown in Fig.~\ref{fig_amr}.

We now have excellent quality data for resolved AGB stars in regions
with well-characterized SFH, spanning a wide range in metallicity,
such as the metal-rich fields of M31 from the  
Panchromatic Hubble Andromeda Treasury (PHAT) survey, which also
includes more than 500 star clusters 
\citep[][see Girardi's contribution, this conference]{Dalcanton_etal12, Johnson_etal12}, 
metal-poor dwarf galaxies up to a distance of 4 Mpc from the
ACS Nearby Galaxy Survey Treasury (ANGST) survey and other 
dwarf galaxies of the Local Group (see Rosenfield's, Boyer's, and Menzie's 
contributions, this conference), 
a complete census of AGB stars in the Magellanic Clouds  at near and 
intermediate IR wavelengths from 2MASS and Spitzer surveys
\citep[][see also Sloan's contribution, this conference]{NikolaevWeinberg_00, Blum_etal06, Bolatto_etal07}, as well as 
for MCs fields with spatially resolved 
SFH derived from the VMC survey \citep[][]{Rubele_etal12}, 
AGB data for the Milky Way (MW) bulge from 2MASS, OGLE and WISE 
\citep{Nikutta_etal14}.
Additional observational constraints are provided by the distributions of
mass-loss rates of AGB stars in the MCs derived from radiative transfer models and
spectral fitting of Spitzer data 
\citep[][see also Srinivasan's contribution, this conference]{Gullieuszik_etal12, Riebel_etal12},
wind expansion velocities, and circumstellar molecular and dust chemistry from
sub-mm and radio observations with Herschel, ALMA, and other telescopes
\citep[][see also Olofsson's and Nanni's contributions, this conference]
{Justtanont_etal12, Schoier_etal13, Nanni_etal14, Nanni_etal13, Vlemmings_etal13, RamstedtOlofsson_14}, 
detailed information about the pulsation
properties of AGB stars provided by the OGLE data 
\citep[][]{Soszynski_etal09, Soszynski_etal13}, and 
stellar parameters from interferometry \citep{Paladini_etal11, Klotz_etal13, vanBelle_etal13}.
\section{A few steps along the calibration cycle}
I will briefly address here two critical quantities the primarily need to be calibrated 
as a function of the stellar progenitor's mass and metallicity,
namely: the TP-AGB lifetimes and the core-mass growth on the TP-AGB.
\subsection{TP-AGB lifetimes}
Assessing the  duration of the TP-AGB, $\tau_{\rm TP-AGB}$, as a function of
the initial stellar mass  is of paramount importance for two main reasons. First,
it controls the energy output of a TP-AGB star,  
$E_{\rm TP-AGB}\propto \int_0^{\tau_{\rm TP-AGB}} L(t) dt$, 
hence its contribution to 
the integrated light of the host system. Second, 
it regulates the number of thermal pulses suffered during the phase, 
hence the degree of surface 
chemical enrichment due to the mixing episodes, ultimately affecting the 
ejecta expelled into the interstellar medium.

\begin{figure*}
\begin{minipage}{0.49\hsize}
\resizebox{1.0\hsize}{!}{\includegraphics{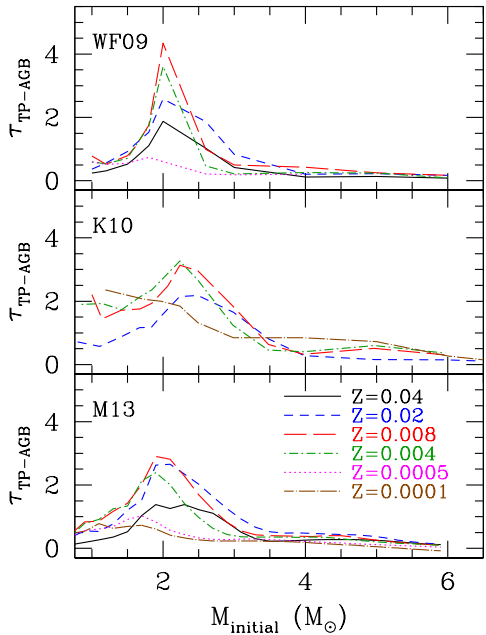}}
\end{minipage}
\hfill
\begin{minipage}{0.48\hsize}
\caption{TP-AGB lifetimes as a function of the stellar initial mass and metallicity.
Predictions of three recent studies are shown, from top to bottom: 
\citet[][WF09]{WeissFerguson_09}, 
\citet[][K10]{Karakas_10}, \citet[][M13]{Marigo_etal13}.} 
\label{fig_tauall}
\end{minipage}
\end{figure*}

While in older TP-AGB models large differences in $\tau_{\rm TP-AGB}$
existed among different studies, a closer agreement is found in more
recent works (Fig.~\ref{fig_tauall}).  All predicted relations between
$\tau_{\rm TP-AGB}$ and the initial stellar mass display a prominent
peak at of $M_{\rm initial}\approx 2 M_{\odot}$, that takes place in
the proximity of the critical stellar mass for the development of a
degenerate He-core at the end of the main sequence, and corresponds to
the minimum in the core mass at the onset of thermal pulses.  We note 
that the dependence of $\tau_{\rm TP-AGB}$ on the
initial metallicity is not monotonic, as it is the result of the
interplay among many different factors (e.g., core
mass at the first TP, evolution of the surface C/O, mass loss, etc.).

In general, the main parameter that directly governs TP-AGB lifetimes 
is the efficiency of mass loss, which may vary with the metallicity.
In the low-metallicity regime, not
covered by the MCs,  valuable constraints on $\tau_{\rm TP-AGB}$ 
are provided by the nearby dwarf galaxies  observed with ANGST \citep{Dalcanton_etal09}.

Following the analysis carried out by \citet{Girardi_etal10}, 
and more recently by \citet[][see Rosenfield's contribution, this conference]{Rosenfield_etal14}, 
it turns out that in order to reproduce the optical and nea-IR star counts and luminosity functions 
of low-mass low-metallicity TP-AGB, it is necessary that significant mass loss 
rates ($\dot M \approx 10^{-7}-10^{-6}\, M_{\odot}$  yr$^{-1}$) are 
already attained 
at rather low luminosities, before the onset of large-amplitude pulsation, 
when dust is not expected to be the dominant driver of stellar winds.
A suitable mode might be provided by the 
flux of Alfv\'en wave  energy associated to cool chromospheres 
\citep{SchroederCuntz_05}.
In this framework novel theoretical efforts \citep{CranmerSaar_11}
indicate that the mass-loss rate should have a steeper, rather than linear, 
dependence  on the magnetic flux, which results into a higher mass-loss efficiency 
at lower metallicity.
\begin{figure*}
\begin{center}
\begin{minipage}{0.47\hsize}
\resizebox{0.8\hsize}{!}{\includegraphics{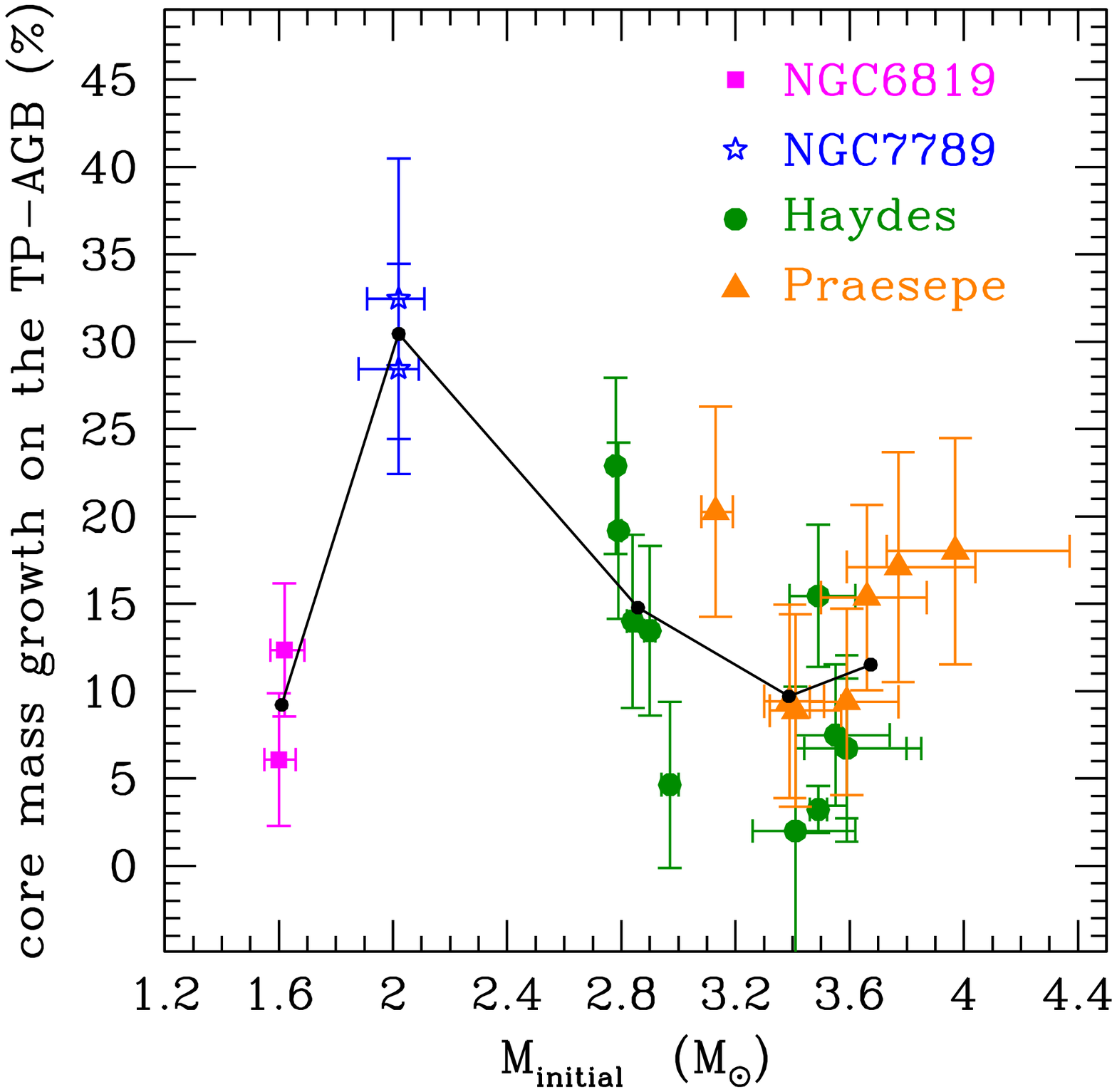}}
\end{minipage}
\hfill
\begin{minipage}{0.47\hsize}
\resizebox{0.8\hsize}{!}{\includegraphics{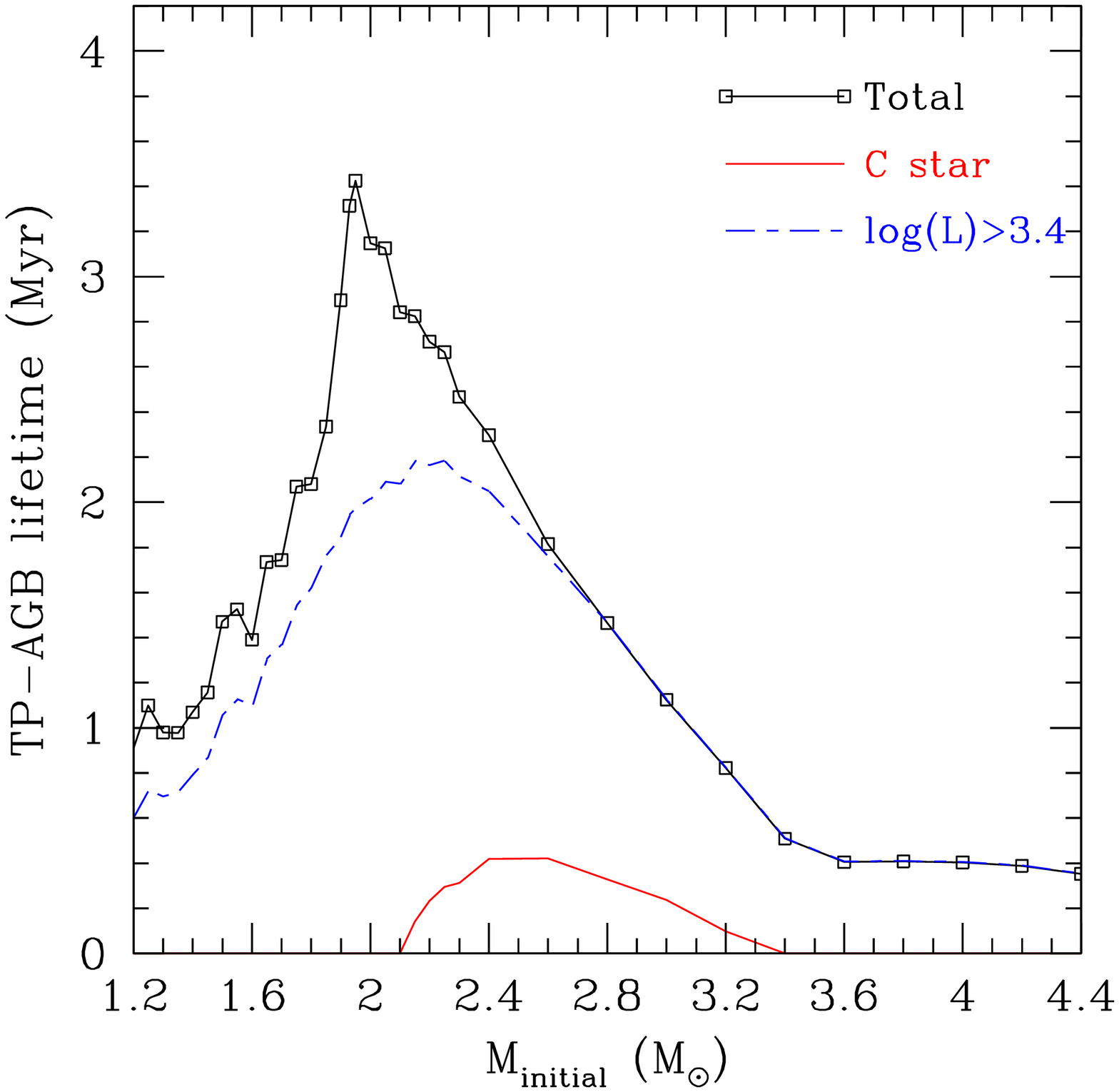}}
\end{minipage}
\caption{{\em Left}: Core-mass growth on the
TP-AGB   relative to new homogeneous, high signal-to-noise WD mass data  
in intermediate-age open clusters \citep{Kalirai_etal14}. 
{\em Right}: The TP-AGB lifetimes from 
 \citet{Marigo_etal13} models with initial metallicity $Z= 0.02$.
The predicted TP-AGB core-mass growth in these models fits the new 
\citep{Kalirai_etal14} measurements very nicely.}
\label{fig_ifmr}
\end{center}
\end{figure*}
\subsection{The core-mass growth on the TP-AGB}

Another parameter of high importance is the core mass growth on the TP-AGB, 
as it determines the amount of mass of the chemically enriched gas 
that is returned to the ISM, and fixes a lower limit to the nuclear fuel
burnt ($\propto$ emitted light) during the TP-AGB phase.
The remaining part of the TP-AGB fuel is expelled 
in the form of chemical yields \citep{MarigoGirardi_01}.
A direct constraint comes from the semi-empirical $M_{\rm initial}-M_{\rm final}$ 
relation that links the initial mass of the progenitor to that of the white dwarf (WD).
The Galactic data shows a clear positive correlation, but with a large scatter, 
which has so far prevented us from obtaining a stringent constraint on AGB models 
\citep[see][for a recent review]{Marigo_13}, though some excellent work has been recently 
carried out on the low-mass end using WDs in old open clusters 
and in common-proper-motion pairs
\citep{Kalirai_etal08, Catalan_etal08}.

Theory predicts that the final mass depends on the core mass growth during the TP-AGB, 
which is affected by two main processes: mass loss and third dredge-up.
The final mass of a TP-AGB is expected to be lower at increasing efficiency of both processes:
while the strength of stellar winds directly controls the overall duration of the TP-AGB phase,
the third dredge-up reduces the net increase of the core mass.
Unfortunately, the efficiencies of both processes are quite uncertain on the theoretical grounds
and need to be observationally constrained.

New constraints on the intermediate-mass range of the $M_{\rm initial}-M_{\rm final}$ 
relation at $Z\simeq 0.02$ have been recently established from newly discovered WDs 
in the nearby Hyades and Praesepe star clusters \citep{Kalirai_etal14},  
and from an accurate re-analysis  of the WD data in the older NGC 6819 and NGC 7789
star clusters, covering a range of progenitors' masses, 
$1.6\, M_{\odot} \la M_{\rm initial} \la  3.8\, M_{\odot}$, over 
which TP-AGB stellar models predict the maximum core growth.
\begin{figure}
\begin{center}
\begin{minipage}{0.48\hsize}
\resizebox{\hsize}{!}{\includegraphics{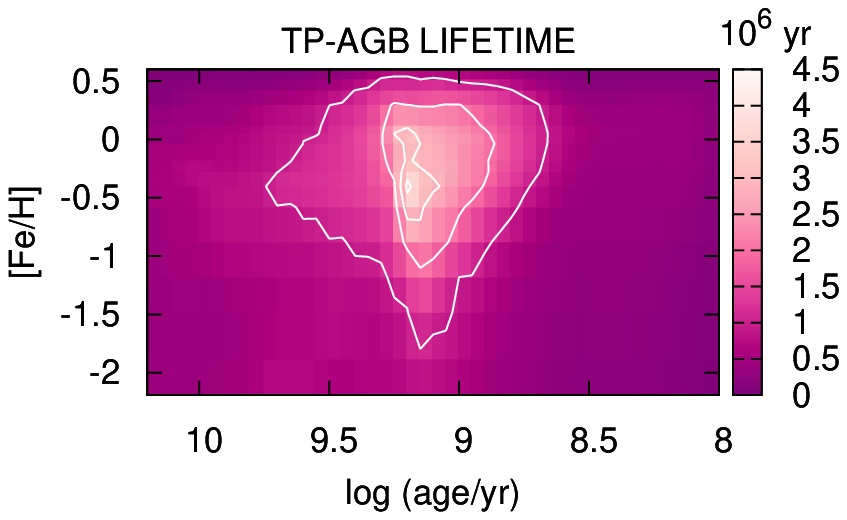}}
\end{minipage}
\hfill
\begin{minipage}{0.48\hsize}
\resizebox{\hsize}{!}{\includegraphics{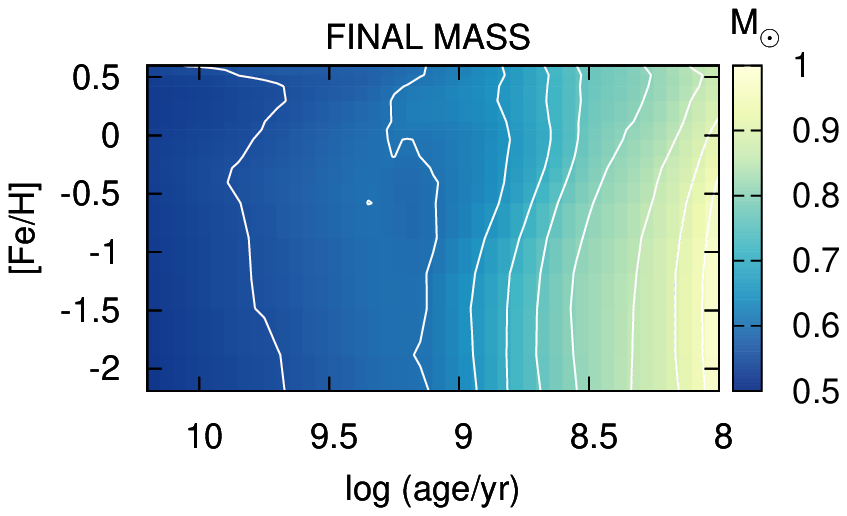}}
\end{minipage}
\caption{Maps of TP-AGB lifetimes ({\em left}) and final core masses ({\em right})
over the relevant age-metallicity plane occupied by TP-AGB stars. 
They derive from 
the TP-AGB calculations of \citet{Marigo_etal13} with the new prescriptions
for pre-dust mass loss as in \citet{Rosenfield_etal14}, and the indications from 
the new $M_{\rm initial}-M_{\rm final}$ relation of \citet{Kalirai_etal14}.}
\label{fig_maps}
\end{center}
\end{figure}

Using the new WD data to calibrate the efficiencies of 
mass loss and third dredge-up in TP-AGB models \citep{Marigo_etal13, MarigoAringer_09},   
we derive that at $Z\simeq 0.02$ i) the carbon star formation should 
be little efficient, 
taking place only in stars with $2.1 \,M_{\odot} \la M_{\rm initial} \la 3.4 \, M_{\odot}$ 
(see Fig.~\ref{fig_ifmr}), which is in line with recent findings for metal-rich
fields of M31 \citep{Boyer_etal13}; ii) 
the peak of TP-AGB lifetimes reaches $\simeq 2$ Myr 
at $M_{\rm initial}\simeq 2.0 M_{\odot}$ for luminosities brighter than the RGB tip; 
iii) the  integrated emitted 
light  peaks at $M_{\rm initial}\simeq 2.4 M_{\odot}$ with 
$E_{\rm TP-AGB} \sim 12 \times 10^{10}$ L$_{\odot}$ yr.
We note that the corresponding  TP-AGB nuclear fuel $(\simeq 0.13 M_{\odot})$ 
is lower by up a factor of 2 than 
previous estimates for $Z=0.02$ \citep{Maraston_05, Marigo_etal08},
supporting other independent claims towards a lighter TP-AGB (see Sect.~\ref{sect_intro}).
\section{The island of TP-AGB stars}
Combining together the results of i) previous studies dealing with TP-AGB stars and WDs in
the MCs and the Galaxy 
\citep{MarigoGirardi_07, WeissFerguson_09, Bianchi_etal11, Kamath_etal12}, 
ii) more recent works based on the ANGST galaxies in the low-metallicity regime
\citep{Girardi_etal10, Rosenfield_etal14}, and iii) the
new constraints on the initial-final mass relation at slightly super-solar metallicity
\citep{Kalirai_etal14}, we can derive a preliminary map of the TP-AGB calibration
over the age-metallicity plane, 
in terms of lifetimes and final core masses (Fig.~\ref{fig_maps}).
A striking result is that there seems to be a special region in the age-metallicity plane, 
a sort of island, which offers the best conditions for the development of TP-AGB phase.
This fertile place encompasses the metallicities going from solar-like values 
to those  typical of the MCs, while the favourite  ages ($\sim$ few Gyr) correspond
to initial stellar masses in the range $\sim 1.5-2.5\, M_{\odot}$. 
Of course, this finding needs to be tested and refined further,  
exploiting the plenty of observed data for resolved TP-AGB stars at our 
disposal today (Sect.~\ref{sect_obs}). 
Work is in progress.

\acknowledgements {This research is supported under ERC Consolidator Grant funding scheme 
(project STARKEY). Special thanks go to the STARKEY team, in particular: B. Aringer, 
S. Bladh, A. Bressan, J.J. Dalcanton, L. Girardi, M.A.T. Groenewegen,  S. H\"ofner, 
T. Lebzelter,  J. Montalb{\'a}n, A. Nanni, P. Rosenfield, M. Trabucchi, P.R. Wood. 
It is a pleasure to thank the ``Vienna AGB working group'', M. Boyer, J. Cannon,  
P. Goudfrooij, M. Gullieuszik, J. Kalirai, C. Paladini, and S. Rubele for fruitful collaboration; 
G. Bruzual, S. Charlot,  and L. Bianchi for valuable feedback and support;
all SOC and LOC members for this exciting conference in the charming Vienna!}
\bibliography{marigo}

\end{document}